# Martin-Löf *à la* Coq

Arthur Adjedj
ENS Paris Saclay, Université
Paris-Saclay
Gif-sur-Yvette, France

Meven Lennon-Bertrand
University of Cambridge
Cambridge, United Kingdom

Kenji Maillard
Inria
Nantes, France

Pierre-Marie Pédrot
Inria
Nantes, France

Loïc Pujet
University of Stockholm
Stockholm, Sweden

**Abstract**

We present an extensive mechanization of the metatheory of Martin-Löf Type Theory (MLTT) in the Coq proof assistant. Our development builds on pre-existing work in Agda to show not only the decidability of conversion, but also the decidability of type checking, using an approach guided by bidirectional type checking. From our proof of decidability, we obtain a certified and executable type checker for a full-fledged version of MLTT with support for $\Pi$, $\Sigma$, $\mathbb{N}$, and **Id** types, and one universe. Our development does not rely on impredicativity, induction-recursion or any axiom beyond MLTT extended with indexed inductive types and a handful of predicative universes, thus narrowing the gap between the object theory and the metatheory to a mere difference in universes. Furthermore, our formalization choices are geared towards a modular development that relies on Coq's features, *e.g.* universe polymorphism and metaprogramming with tactics.

***Keywords:*** Dependent type system, Bidirectional typing, Logical relations

## 1 Introduction

Self-certification of proof assistants is a long-standing and very enticing goal. Since proof assistant kernels are by construction relatively small, have a precise specification, and are part of the trusted computing base of any software certified using them, they offer a natural target for certification. Yet, full certification of a realistic kernel based on dependent type theory remains a challenging goal.

The reasons are twofold. First, dependent type theory is very expressive, allowing to formulate mathematical statements with small amounts of encoding. Consequently, even a minimalistic kernel has to be rather complex. Second, and more critically, dependent type theory intertwines types and computations, forcing us to prove results about computations in order to fully certify a type checker. The usual approach is to establish a *normalization* result and then derive the decidability of the conversion relation, which is used to compare types during type checking. Normalization, however, is generally difficult to prove, as it typically implies soundness of the type system seen as a logic. Accordingly, an important part of the work needed to fully certify a type checker is spent on establishing meta-theoretic properties, which are necessary to ensure termination of the type checker but have little to do with its concrete implementation.

Acknowledging this tension leads to two radically different approaches. On the one hand, one can simply postulate normalization, to better concentrate on the difficulties faced when certifying a realistic type-checker. The most ambitious project to date that follows this approach is Meta-Coq [Sozeau, Anand, et al. 2020; Sozeau, Forster, et al. 2023], which formalizes a nearly complete fragment of Coq's type system and provides a certified type checker aiming for execution in a realistic context, after extraction. On the other hand, one can concentrate on normalization and decidability of conversion, which are the most difficult theoretical problems. The most advanced formalizations on that end are Abel, Öhman, et al. [2017] and Wieczorek and Biernacki [2018]. The first, in Agda, shows decidability of conversion, but does not provide an executable conversion checker. The second, in Coq, certifies a conversion checker designed for execution after extraction, but supports a type theory that is less powerful than the former, *e.g.* it does not feature large elimination of inductive types. Neither formalization provide a type checker.

***Contributions.*** We aim to bring together and improve on this state of the art on proof assistant certification. More precisely, we:

- formalize a full-fledged version of Martin-Löf type theory (MLTT) that features dependent function ($\Pi$) and pair ($\Sigma$) types with $\eta$-laws, natural numbers $\mathbb{N}$ and intensional equality **Id** types with large elimination, and one predicative universe;
- show decidability of both conversion and type checking, refining the logical relation of Abel, Öhman, et al. [2017];
- provide a naïve but certified and executable implementation of the type checker;
- define our model in MLTT with indexed inductive types (*i.e.* we do not rely on either induction-recursion or impredicativity), which provides a narrow upper bound on the logical strength needed to show normalization;

- design our development in a modular way that accommodates extensions of the type theory.

The ongoing development is freely accessible on Github [Adjedj et al. 2023a], and the version referenced in this article is available on Zenodo [Adjedj et al. 2023b]. Our formalization spans around 30k lines of code, and we refer to it in the text with links in blue. To help navigate it, a (file-level) dependency graph is given in Appendix A.

***Plan of the paper.*** Section 2 sets the context of this work, giving a high-level tour of the theory and the metatheoretical properties we formalize. We provide a detailed comparison with prior work in Section 3. Section 4 presents the logical relation technique and Section 5 details some challenges of our formalization to encode logical relations for MLTT and establish a bound on the complexity of the normalization proof. Section 6 explains the algorithmic aspects of typing and conversion, culminating with decidability. Section 7 details engineering aspects of the formalization, while Section 8 explore the future opportunities created by this work.

## 2 MLTT and its metatheory

We take Martin-Löf type theory [Martin-Löf and Sambin 1984]—or rather the Martin-Löf logical framework, which encompasses a whole family of type systems—as an ideal version of the actual type system implemented by proof assistants based on dependent types, such as AGDA, COQ or LEAN. MLTT is presented in terms of judgements for well-formed contexts and types (written $\vdash \Gamma$ and $\Gamma \vdash T$), well-typed terms ($\Gamma \vdash t : T$), but also *conversion* judgements for types ($\Gamma \vdash T \cong T'$), and terms ($\Gamma \vdash t \cong t' : T$) which assert that the two sides are definitionally equal, and can be used interchangeably. Importantly, conversion is typed: it only makes sense to compare terms *at a given type*. Judgements are derived according to typing rules that we leave out to the formalization, see DeclarativeTyping. In our development, we implement in this logical framework the following type formers:

- the dependent function types $\Pi x \colon A.B$ (with a definitional $\eta$-rule);
- the dependent pair types $\Sigma x \colon A.B$ (eliminated through projections, with a definitional $\eta$-rule);
- a universe of small types, noted Type;
- the inductive datatype of natural numbers $\mathbf{N}$, with an induction principle that allows large elimination;
- the Martin-Löf identity type $\mathrm{Id}_A\ x\ y$, and the J eliminator that supports large elimination;
- the empty type $\mathbf{0}$ and its eliminator.

Our goal is to give a formal proof of the metatheoretical properties that make Martin-Löf type theory a well-behaved foundation for proof assistants: consistency, canonicity and decidability. *Consistency* states that there is no inhabitant of the empty type in the empty context, *i.e.* that the logic is sound. *Canonicity* asserts that every inhabitant of $\mathbf{N}$ in an empty context is convertible to a numeral, *i.e.* a succession of successors applied to $0$. Finally, typing and conversion should be *decidable*, more precisely it should be decidable whether any typing or conversion judgement is derivable.

Establishing these properties, however, is challenging. In particular, *large elimination*, *i.e.* the possibility to construct a type by induction, deeply intertwines the term and type levels. This fundamental feature of inductive types in MLTT sets this type theory apart, giving it its expressivity but complexifying its metatheory. In Section 4, we present a logical relation based on *reduction* to tackle this problem.

Indeed, conversion can be presented as the symmetric, reflexive, transitive, congruence closure of a reduction relation, modulo additional rules such as $\eta$-laws. Our formalization uses *weak-head reduction* $\leadsto$, a deterministic[1] reduction strategy that only reduces redexes at the top of a term but not in subterms. The normal forms for this reduction strategy, called weak-head normal forms, can be characterized inductively as either a canonical introduction form for a type (e.g. $\lambda x \colon A.t$ for $\Pi$, $0$ or $S\ n$ for $\mathbf{N}$), or a *neutral* term, that is a stack of elimination forms (e.g. application $n\ t$ for $\Pi$, or induction $\mathrm{ind}_{\mathbf{N}}(n; P; t)$ for $\mathbf{N}$ for neutrals $n$) ultimately stuck on a variable. The logical relation will be used to prove two key properties: first, *weak-head normalization*, which asserts that all well-typed terms reduce to a weak-head normal form; second, *subject reduction*, which stipulates that types are invariant by reduction (if $x : A$ and $x \leadsto y$ then $y : A$). From these, consistency of the system can be directly derived. The algorithmic presentations of typing and conversion from Section 6 also heavily rely on these properties to prove canonicity and decidability.

## 3 Related work

In response to the challenge of establishing metatheoretical properties for MLTT, the literature can be divided into two trends. On the one hand, there is a natural incentive to leverage the power of proof assistants to keep track of much of the nitty-gritty details and even automate them away. On the other hand, there is a growing interest in sophisticated frameworks that abstract away from these details in order to give synthetic, although mathematically challenging proofs.

***Towards certified proof-assistants.*** The METACOQ project [Sozeau, Anand, et al. 2020; Sozeau, Forster, et al. 2023] provides a certified type checker for a type system very close to the one underlying COQ, assuming normalization of the formalized type theory. It is based on an untyped presentation of conversion, which facilitates parts of the metatheoretical work, but makes it much more difficult to handle extensionality rules, *e.g.* the $\eta$ rules for functions and primitive records, which are currently not supported.

---

[1]For any $t$, there exists at most one $t'$ such that $t \leadsto t'$.

Much earlier, Barras and Werner [1997] provide a fully certified proof assistant for the Calculus of Constructions, even featuring a small REPL on top of the certified kernel.

Outside dependently typed proof assistants, very advanced certification efforts have been achieved for the HOL family with the Candle project [Abrahamsson et al. 2022].

***Formalized meta-theory of dependent type systems.*** In the late '90s, Barras and Werner [1997] fully certified a proof assistant for the Calculus of Constructions (CC). They proved the normalization and decidability of type checking for CC in Coq using reducibility candidates [Girard et al. 1989]. Their type theory supports an impredicative universe, but no inductive types. This allows Barras *et al.* to erase dependency from the types in their normalization proof, and thereby solve the interdependency puzzle without too much difficulty. In his PhD thesis, Barras [1999] adds inductive types and a hierarchy of predicative universes, and shows decidability of type checking assuming normalization, but to the best of our knowledge this part of his work was never formalized. In later unpublished work, Barras [2014] shows normalization of CC extended with natural numbers and large elimination, as well as consistency of the Extended Calculus of Constructions, adding a countable hierarchy of predicative universes to CC, together with W-types. His formal proof relies on realizability models built on top of an embedding of IZF in Coq, and has been applied to establish the metatheory of CoqMT [Wang and Barras 2013].

More recently, Wieczorek and Biernacki [2018] formally prove the correctness of a normalization by evaluation (NbE) algorithm in Coq, following pen-and-paper proofs by Abel and co-authors [Abel, Coquand, and Dybjer 2007; Abel, Aehlig, et al. 2007; Abel, Coquand, and Pagano 2009; Abel 2010]. Compared to the work of Barras *et al.*, their development covers some primitive datatypes, namely natural numbers, and the empty and unit types. However, it does not tackle large elimination, the main difficulty presented by inductive types in this setting. For their normalization proof, Wieczorek *et al.* model types as proof-irrelevant partial equivalence relations (PERs). Since the pen-and-paper definition uses induction-recursion, which is not available in Coq, the authors replace it with impredicative encodings, creating in important gap in logical power between the object theory and their meta-theory. They provide a conversion checker designed to be run after extraction (which erases the complex termination argument), but not a type checker, which they leave as future work.

In parallel, Abel, Öhman, et al. [2017] attack a similar problem in Agda. Rather than proving correctness of normalization by evaluation in a semantic domain, they rely on a reduction-based approach, although they nevertheless use a typed notion of conversion which supports $\eta$ laws. Moreover, they give a proper treatment of large elimination for the inductive type of natural numbers and its induction principle. The definition of their logical relation is also based on the work of Abel [2010] *etc.*, but it relies on induction-recursion instead of impredicativity. Similarly to Wieczorek *et al.*, they stop after decidability of conversion, and their decider is moreover not geared toward execution, neither in Agda nor after extraction.

This work has since been extended to justify multiple additions to MLTT [Gilbert et al. 2019; Pujet and Tabareau 2022, 2023; Abel, Danielsson, et al. 2023]. Of particular interest is Pujet and Tabareau [2023], which lowers the logical power needed for the proof by removing induction-recursion with an encoding similar to ours. Yet, this proof still relies on Agda's first-class universe levels, whose meta-theory has been relatively unexplored, but for Kovács [2022].

In his lecture at the Collège de France, Leroy [2020] reviews these recent approaches to the certification of proof assistants, in particular for Coq and Agda, emphasizing the need for sharper consistency bounds as we provide here. More broadly, the two POPLMark challenges [Aydemir et al. 2005; Abel, Allais, et al. 2019], which aim at enhancing the general work on formalized meta-theory of type systems, are also relevant to this work. Indeed, the AutoSubst plugin for Coq [Stark et al. 2019], which has been developed in this setting, has been directly useful to us, and further developments on modular mechanization of metatheory [Jin et al. 2023; Delaware et al. 2013; Forster and Stark 2020] would likely make developments such as ours much easier.

***Frameworks for the meta-theory of dependent types.*** In recent years, the community has shown a renewed interest in higher-level proofs that pack tedious details into convenient abstractions, and make more room for the big picture. Such proofs have been formulated in the categorical language of gluing [Coquand 2018; Bocquet et al. 2023], as well as more sophisticated frameworks such as synthetic Tait computability [Sterling 2021; Gratzer et al. 2019]. Abstraction aside, the features that set these apart from more traditional proofs are the use of proof-relevant logical relations on the one hand, and the replacement of partial equivalence relations with actual quotients on the other hand. Accordingly, proofs of normalization by gluing have a very extensional flavor, and as such are less amenable to implementation in a proof assistant based on intensional type theory. Moreover, the more sophisticated iterations rely on (multi)modal type theories as internal languages for feature-rich categories, for which mechanization is still in its infancy. Finally, if the goal is to obtain a certified implementation, the connection between these abstract proofs and an executable algorithm has yet to be explicited.

***Partiality and general recursion in type theory.*** Conversion checking algorithms are not structurally recursive in their inputs, and indeed their termination argument is highly complex, since it relies on normalization. Implementing a conversion checker in a type theory admitting only structural recursion, such as Coq's, is thus challenging. The more traditional approaches to non-structurally recursive functions either rely on well-foundedness, encoded as an inductive accessibility predicate; or on step-indexing, using an extra "fuel" parameter bounding the allowed number of recursive calls. The latter induces significant noise in the definition, while the former makes it impossible to separate the definition of a function from a proof of its termination/totality, and impedes computation by making it necessary to compute very large accessibility proofs. Alternatives include using co-inductive types [Capretta 2005], and the Bove and Capretta [2005] approach, relying on an inductive characterization of the function's graph. A variant of the latter is the Braga method [Larchey-Wendling and Monin 2022], which adds the extra goal of a well-behaved extraction. All of these allow to define partial functions, without the extra complexity of carrying a fuel parameter around. Termination (possibly only on a subset of the domain) can be established separately from the function definition.

Our approach, based on [McBride 2015], separates the description of a recursive function from its realization using a free monad to describe the calling graph. From this monadic object, the fuelled, coinductive or graph-based realization can all be easily recovered, depending on one's goals. As with the Bove-Capretta and Braga methods, this lets us define functions before arguing about their termination, but without foregoing the ease of execution of the fuelled variant. This part of our development rely on a library by Winterhalter [2023], which implements an enhanced version of McBride's ideas in Coq.

## 4 The logical relation(s)

We establish the metatheoretical properties from Section 2 through an equivalence between two presentations of MLTT, a declarative presentation in the tradition of Martin-Löf's logical framework and an algorithmic one based on ideas from bidirectional typing (cf Section 6). The latter is used as an effective specification to prove the correction of the type checking algorithm. In this section, we give a high level tour of the logical relation techniques that we employ to relate both presentations. We roughly follow Abel, Öhman, et al. [2017], where a more detailed account can be found.

### 4.1 Metatheory through logical relations

The point of the logical relation is to build a model of dependent type theory where types are interpreted as *reducibility* predicates, *i.e.* predicates on the untyped syntax that characterize well-behaved inhabitants. In order to model convertibility, these reducibility predicates are additionally equipped with a relation that we call *reducible conversion*, effectively turning them into PERs on untyped terms.

The first step in the construction of our model is thus the mutual definition of a reducibility predicate to characterize types (type-level reducibility), together with a reducibility predicate to characterize the inhabitants of every reducible type (term-level reducibility), and the associated reducible conversions. Then, we show that reducible conversions are equivalence relations (reflexive, symmetric, transitive), that reducibility implies weak-head normalization, is closed under weakening and anti-reduction, and is irrelevant, meaning that the term-level reducibility and reducible conversions associated to reducibly convertible types are equivalent. Next, in order to interpret the typing judgements in our model, we need to give a semantic interpretation of contexts and substitutions. To this end, we define the validity relation by (freely) closing reducibility under substitutions of reducible terms. Finally, we prove the fundamental lemma by induction on the typing derivations, which states that any derivable declarative judgement is valid. Once the fundamental lemma is established, we obtain that any well-typed term has a weak-head normal form classified by (the weak-head normal form of) its type, from which we derive many metatheoretical properties: injectivity of type constructors (hence subject reduction), consistency and canonicity.

### 4.2 Three logical relations in one

The fundamental lemma is enough to establish the normalization properties but proving decidability of the conversion and of the type-checking requires more work. We first prove that the algorithmic presentation is complete with respect to derivability in the declarative system: any judgements derivable in the declarative system is also derivable in the algorithmic system. The completeness proof uses another logical relation defined in terms of the algorithmic system instead of the declarative one. In order to factor out the work, we reuse an idea of Abel, Öhman, et al. [2017] who parameterize the logical relation by an abstract conversion relation. We expand over this idea and parameterize the definition of the logical relation by a generic typing interface accommodating both declarative and algorithmic typing.

A first instantiation of the interface with declarative typing gives us among other (weak-head) normalization and injectivity of type constructors. Two more instantiations, one with a mixed system combining declarative typing and algorithmic conversion, and one with a fully algorithmic system, let us relate the declarative and algorithmic judgements (see Section 6.4).

### 4.3 Design choices for the logical relation

Our model relies on an untyped small-step reduction. That is, even though at times we bundle big-step reduction proofs with side-conditions that one or both sides are well-typed, we do not ask for typing proofs at the granularity of single reduction steps. As a result, subject reduction is a result that becomes available late, after the fundamental lemma has been proven.

Since this goes against the position explicitly advocated for in [Abel, Öhman, et al. 2017], we believe that this design choice deserves some discussion. The first reason is purely rooted in engineering considerations. Basically, asking for a typed reduction duplicates the definition of typing derivations into a reduction variant. These inductive types are big, so they require a lot of boilerplate. It also makes experimenting with extensions to the type theory cumbersome, since one has to duplicate the additions in both inductive types. Given that typing itself is already quite redundant with conversion, this was deemed too unpractical.

Another more theoretical reason is that opting for typed reduction hardwires a specific kind of models, or equivalently a specific kind of generic typing interfaces. Since typed reduction implies typing in the declarative system, and the logical relation implies reduction to a normal form, it is essentially asking that the resulting model is complete for declarative typing. Nonetheless, there are instances of typing interfaces that either cannot be proven complete early (i.e. before the fundamental lemma), or simply are not complete. A typical example of the latter is the instance used to prove untyped weak-head normalization of well-typed terms, which interprets all typing statements trivially.

### 4.4 Abstract conversion of neutrals

While reducible conversion finely characterizes the behaviour of canonical terms, it is essentially blind to the structure of neutral terms. Indeed, two neutrals are reducibly convertible simply if they are related by an abstract notion of neutral conversion, which is part of the generic typing interface. While we can recover properties of neutrals by a well-chosen instantiation of the interface, due to this structure neutral conversion cannot be defined mutually with reducibility, limiting its power. Hence, the declarative instance of the logical relation does *not* show that neutral destructors are injective (e.g. that if $\Gamma \vdash n\ u \cong n'\ u' : T$ with $n$ and $n'$ neutral, then $n$ and $u$ are respectively convertible to $n'$ and $u'$). This is the core reason why we must instantiate the logical relation with algorithmic instances, to obtain properties of algorithmic neutral comparison which we cannot get more directly. The declarative instance also does not imply deep normalization, again because it does not go under neutrals. This is why the algorithm's proof of termination (Section 7.2) works directly on a conversion derivation.

## 5 Not-so-small induction-recursion

The definitions of reducibility and validity that we outlined in Section 4.1 are challenging to express in type theory. In the case of reducibility, term-level reducibility is indexed over type-level reducibility, but type-level reducibility must depend on term-level reducibility in order to properly model dependent types that contain terms. Likewise, validity for contexts is mutually defined with validity for types and substitutions. In presence of inductive types with large elimination, these dependencies cannot be swept under the rug and must be taken into account in our model.

In their proof, Abel, Öhman, et al. [2017] solve this dependency puzzle by exploiting the powerful definition scheme of Agda, which allows to mutually define an inductive type with a function defined by recursion on that very type. This feature is known as *induction-recursion* (IR for short) [Dybjer and Setzer 2003], and is commonly used to build models of dependent type theory. Thus, Abel *et al.* use IR to simultaneously define reducible types $\Gamma \Vdash \langle \ell \rangle\ A$ in context $\Gamma$ by induction, and associate an adequate reducibility predicate $\Gamma \Vdash \langle \ell \rangle\ t : A\ /\ \mathfrak{R}_A$ to every reducibility proof $\mathfrak{R}_A : \Gamma \Vdash \langle \ell \rangle\ A$, by recursion on $\mathfrak{R}_A$.

For instance, there is a constructor $\mathfrak{red}_\mathbb{N}$ that, given any type $A$ and a proof $r : A \rightsquigarrow^* \mathbf{N}$ that it weak-head reduces to $\mathbf{N}$, induces a proof $\Gamma \Vdash \langle \ell \rangle\ A$. Correspondingly, the recursive case $\Gamma \Vdash \langle \ell \rangle\ t : A\ /\ \mathfrak{red}_\mathbb{N}\ A\ r$ is defined as reducibility at the type of natural numbers $\Gamma \Vdash_\mathbb{N} t$. The latter asserts that the term $t$ is reducible if it weak-head reduces either to 0, to a successor $\mathsf{S}\ u$ with $\Gamma \Vdash_\mathbb{N} u$ reducible at $\mathbf{N}$ too, or to a neutral term of type $\mathbf{N}$ in context $\Gamma$. Note that the recursive definition of the reducibility predicates on terms is already needed in the simply typed setting to account for function types $A \rightarrow B$ that feature a negative occurrence of reducibility for terms in the domain.

Furthermore, reducibility is indexed by an integer $\ell$ that reflects the stratification of types into universe levels: the definition for $\ell = 0$ only accounts for *small* types and their inhabitants (*i.e.* types that do not mention any universe), while the definition for $\ell = 1$ accounts for all types, large or small. This way, we can declare that a term is a reducible inhabitant of the universe $\Gamma \Vdash \langle 1 \rangle\ t : A\ /\ \mathfrak{red}_\mathsf{U}\ A$ precisely when it is a reducible small type $\Gamma \Vdash \langle 0 \rangle\ t$, without introducing any circularity.

In exchange for its elegance, IR introduces a serious gap between the metatheory and the object theory, which only supports a handful of inductive types. Even though exploring normalization proofs for IR is a valuable endeavour, we would rather go the other way and narrow this gap by recasting the definitions of Abel *et al.* to rely only on regular indexed inductive types. In fact, this restriction is enforced by our choice of proof assistant, since Coq does not support induction-recursion. In exchange, Coq supports impredicativity through its sort of propositions `Prop`, but we never rely on impredicativity in our development[2].

***Removing induction-recursion.*** Although IR is strictly stronger than plain indexed inductive types in general, it

---

[2]We cannot completely avoid `Prop`, due to its ubiquity in Coq's standard library, for instance via the identity type eq. But the proofs would work just as well with a `Type`-valued equality.

```
Definition RedRel@{i} :=
  Con → Term → (Term → Type@{i}) → Type@{i+1}.

Inductive LR@{i} : ∀ (ℓ : TypeLevel), RedRel@{i} :=
| redU : Γ ⊩U A →
  LR@{i+1} 1 Γ A (fun B ⇒ ∑ P, LR@{i} 0 Γ B P)
| redℕ : Γ ⊩ℕ A → LR@{i} ℓ Γ A Redℕ.

Notation "Γ ⊩⟨ℓ⟩ A" := (∑ P, LR ℓ Γ A P).
Notation "Γ ⊩⟨ℓ⟩ t : A / RA" := (fst RA t).
```

**Figure 1.** Simplified excerpt of reducibility via small induction-recursion in Coq

is possible to transform an inductive-recursive definition into an inductive type if the codomain of the recursive part is smaller than the universe of the inductive part—this is known as *small induction-recursion* [Hancock et al. 2013]. The transformation works as follows: we first define the codomain of the recursive part, here the reducibility predicates on term, and then define an inductive predicate carving out those elements of the codomain that arise from the evaluation of the putative inductive part of the induction-recursion instance. The inductive part is then recovered by packing together an element of the codomain together with an inductive proof that it is in the image of the inductive-recursive definition, while the recursive part just projects out the element of the codomain, forgetting the inductive proof.

Reducibility is not *a priori* an instance of small IR, as both the term-level and type-level reducibility are defined in the same universe to enable defining reducibility for inhabitants of the universe in terms of type-level reducibility. However, it turns out that it is possible to recast it as one, at the cost of some universe level juggling: we can define type-level reducibility for small types (the inductive part for $\ell = 0$) in $\mathrm{Type}_1$, while term-level reducibility for small types lands in $\mathrm{Type}_0$. Then, type-level reducibility for large types is defined in $\mathrm{Type}_2$, and the term-level part in $\mathrm{Type}_1$. Proceeding in this fashion, we can define the reducibility predicate for the universe as the type-level reducibility of small types, while keeping the recursive part of the definition smaller than the inductive part.

The result is presented in Fig. 1 in a Coq-inspired syntax. Here, LR@{i} ℓ Γ A P encodes as a functional relation the fact that P : term → Type@{i} is the reducibility predicate $\Gamma \Vdash\langle\ell\rangle \_ : A / \mathfrak{R}_A$ associated to the reducibility proof $\mathfrak{R}_A$ in the usual IR presentation. LR is heavily simplified compared to the development, serving as an illustration device to build up intuition. The first difference is that in addition to term reducibility we also have two additional predicates term → Type@{i} and term → term → Type@{i}, for reducible conversion of types $\Gamma \Vdash\langle\ell\rangle A \cong \_ / \mathfrak{R}_A$ and of terms $\Gamma \Vdash\langle\ell\rangle \_ \cong \_ : A / \mathfrak{R}_A$, respectively.

Furthermore, observe that the universe level of the definition depends on the level $\ell$, which is not possible in Coq since first-class universe levels are not available. Instead, we resort to what amounts to good old-fashioned code duplication, essentially defining reducibility once for small types, then once again for large types. Luckily, we sidestep inconvenient copy-pasting by a mix of universe polymorphism, a form of open recursion, and the definition of a custom induction principle (see Section 7.1 for details).

This stratification is unpleasant, but we cannot hope to eliminate it completely, because of Gödel's incompleteness theorem. Indeed, normalization implies consistency, and we cannot hope to prove consistency for MLTT with arbitrarily many universes in a meta-theory that amounts to MLTT—thus some code manipulation is necessary in order to extend the proof with one additional universe.

Abel *et al.* use induction-recursion a second time to define validity. The same strategy as the one we adopt for reducibility applies to validity as well, and allows us to reformulate it as an indexed inductive type too. As a result, we completely remove induction-recursion, allowing a port of the proof to plain Coq.

***The gap between the object and the meta.*** The meta-theory for MLTT with $n$ universes should ideally be MLTT with $n + 1$ universes, but unfortunately our proof requires at least one additional universe. Indeed, our definition of reducibility for $n$ universes requires $n + 2$ universes in the meta-theory, appearing as $i$ and $i + 1$ in Fig. 1.[3] The definition of validity also fits within these $n + 2$ universes, which means they should be sufficient for the proof to go through. In our development, for an object theory with a single universe, we should be using 3 universes from the meta-theory to define reducibility and validity.

This is the theory, but in practice we assume two additional universe levels that live *below* the universes used for reducibility. These are used in universe polymorphic definitions: since a definition must use the same number of universe variables for $\ell = 0$ and $\ell = 1$, we use these additional universes to instantiate the unneeded variables for the $\ell = 0$ case. In the end, our development uses $n + 4$ universes to show normalization for MLTT with $n$ universes.

## 6 An algorithmic presentation of MLTT

Although decidability of conversion is the main difficulty in a proof that type checking is decidable, the latter does not automatically follow from the former, except for heavily annotated systems – see *e.g.* Petković Komel [2021, Proposition 10.3.5]. In fact, type checking for MLTT as defined

---

[3]Since Coq does not support algebraic universes of the shape $i + 1$ in surface syntax, we need to use two levels $i < j$ in the formalization.

Martin-Löf *à la* Coq

by Abel, Öhman, et al. [2017] is undecidable for this very reason [Dowek 1993; Sørensen and Urzyczyn 2006].

In our development, we cover this last mile and show decidability of typing with a full account of algorithmic type checking, presented in a bidirectional fashion [Pierce and Turner 2000; Dunfield and Krishnaswami 2021]. The main idea of bidirectional typing is to refine typing into type *inference*, where the type of a term is an unknown to be found, and type *checking*, where the type is known, which are mutually defined, both as judgement and as functions. Following the strategy implemented *e.g.* by Coq, our terms contain enough annotations to ensure that inference is complete: *every* well-typed term infers a type. For instance, $\lambda$-abstractions are annotated with the type of the variable, *i.e.* Church-style. We do not investigate the common alternative in the bidirectional setting, which reduces the need for annotations by foregoing completeness of inference – some well-typed terms can only be type checked against a given type, because they do not contain enough type information by themselves.

Of course, algorithmic type checking relies on an algorithmic conversion. Ours is strongly inspired by the one of Abel, Öhman, et al. [2017], but puts more emphasis on its implicit bidirectional character. Indeed, our definition of algorithmic conversion is also decomposed into two mutually defined relations: general conversion checking, which is "checking" in the sense that it takes a type as input in order to compare the two terms; and a special neutral comparison relation, which is "inferring", in the sense that a common type for the two compared terms is synthesized while comparing them.

### 6.1 Algorithmic/bidirectional typing

Bidirectional typing, as an inductive predicate, is defined in AlgorithmicTyping. We denote type inference as $\Gamma \vdash t \triangleright T$, and type checking as $\Gamma \vdash t \triangleleft T$. Each declarative typing rule for a term constructor gives rise to a corresponding rule for type inference, which ensures completeness of inference. For instance, the algorithmic rule for application is

$$\text{INFAPP} \frac{\Gamma \vdash f \triangleright_{\text{h}} \Pi\, x{:}A.B \qquad \Gamma \vdash a \triangleleft A}{\Gamma \vdash f\, u \triangleright B[u]}$$

Since terms might infer a type that does not exactly meet the required constraints, two extra rules let us handle conversion. For instance, in rule INFAPP above, $a$ might not infer $A$, but some $A'$ convertible to $A$. This is exactly what the only rule to derive type checking lets us do:

$$\text{CHECKCONV} \frac{\Gamma \vdash t \triangleright T \qquad \Gamma \vdash T \cong T'}{\Gamma \vdash t \triangleleft T'}$$

However, it can happen that we have a constraint only on the *head constructor* of the inferred type: for instance, the type of first premise of INFAPP should be a $\Pi$ type. In that case, we cannot use the conversion checker to compare the type inferred by $f$ with $\Pi\, x{:}A.B$, as neither $A$ nor $B$ are specified. This typically happens for destructors like application, where the head type constructor of the destructed term is known, but no more. In this case, we use reduction on the inferred type to expose its head constructor, and check that it matches the expected one. This is exactly what *reduced inference*, written $\Gamma \vdash t \triangleright_{\text{h}} T$,[4] does, corresponding to the following rule:

$$\text{INFRED} \frac{\Gamma \vdash t \triangleright T \qquad T \rightsquigarrow^{\star} T'}{\Gamma \vdash t \triangleright_{\text{h}} T'}$$

The choice of using inference, reduced inference or checking in a premise when turning a declarative rule into an algorithmic one is a rather mechanical one. If the type is fully known from earlier premises or the term under consideration, we can use checking. If the type is fully unknown, we must use inference. Finally, when the type is partly unknown but has a prescribed shape we use reduced inference, in order to uncover said shape. In essence, this follows what Dunfield and Krishnaswami [2021] call the "Pfenning recipe", with the extra complication that we rely on reduction to exhibit the shape of types.

### 6.2 Algorithmic conversion is bidirectional too

The easiest way to implement conversion checking is to fully normalize terms to $\eta$-long deep normal forms, and compare these for pure syntactic equality. However, this approach is neither faithful to our logical relation, which proceeds by iterated weak-head normalization, nor to most implementations, which also follow this lazy, stepwise approach. So we instead implement our conversion checker and the inductive relation that presents it by an interleaving of weak-head reduction and structural comparison of head normal forms.

The entry point of algorithmic conversion, in rule CHECK-CONV, is conversion between types, which is fairly straightforward: the two types are reduced to weak-head normal form, and are deemed convertible if they share the same head constructor and all their subtypes/subterms are recursively convertible. However, since these weak-head normal form might be neutral inhabitants of the universe, we must be able to compare arbitrary neutral terms for conversion.

Algorithmic conversion of terms is subtler than for types. Indeed, consider the declarative conversion rules for functions: congruence of application and the $\eta$ rule for functions.

$$\text{TermAppCong} \frac{\Gamma \vdash f \cong g : \Pi\, x{:}A.B \qquad \Gamma \vdash a \cong b : A}{\Gamma \vdash f\, a \cong g\, b : B[a]}$$

$$\text{TermFunExt} \frac{\Gamma, x{:}A \vdash f\, x \cong g\, x : B}{\Gamma \vdash f \cong g : \Pi\, x{:}A.B}$$

How should information propagate in these rules? On one hand, the type $B$ in TermAppCong can only be obtained from the first premise, as we cannot uniquely invert the substitution $B[a]$. On the other hand we would like to use the information that conversion happens at a Π-type to trigger TermFunExt. Thus, type information is useful for type-directed rules such as TermFunExt, yet it is impossible to propagate type information bottom-up through TermApp-Cong. The way out is to split conversion checking in two: a general relation to compare arbitrary terms, which takes a type as input, and can use it to trigger type-directed rules; and a second relation to specifically compare neutral terms, which infers a type, propagating it upside-down rather than bottom-up. We write the former relation $\Gamma \vdash t \cong t' \lhd T$, using the same symbol as for type checking to insist on the bidirectional intuition, and the latter as $\Gamma \vdash n \sim n' \rhd T$. This approach is sensible since extensionality rules are useless on neutrals, intuitively because $\eta$-expanding neutrals is useless as it cannot trigger any further computation, but only add an extra stuck layer.

---
[4] The "h" stands for (weak-)**h**ead reduction.

Once this design decision is fixed, the rest of the definition follows straightforwardly. In the end, conversion checking operates roughly as follows:

1. the two terms to compare and their type are reduced to weak-head normal form,
2. if the type is one with an $\eta$-rule (Π or Σ), then this rule is applied, and the $\eta$-expanded terms are recursively compared,
3. otherwise, if the two terms start with the same canonical constructor, it is stripped and its subterms are recursively compared,
4. finally, if the two terms are neutrals, they are compared using neutral comparison.

Neutral comparison structurally traverses the two neutrals to find the variable on which they are stuck, gets its type from the context, and then uses that type information to recursively compare the other subterms with general conversion. If the variable is not the same, or the neutrals do not have the same structure, they are not convertible.

### 6.3 Bundled algorithmic typing: invariants as an induction principle

There is an important caveat to algorithmic typing as defined in AlgorithmicTyping: it is not, in general, equivalent to declarative typing. The reason is the way we treat "boundaries" of judgements (*e.g.* $\Gamma$ and $T$ in $\Gamma \vdash t : T$). In declarative typing, whenever $\Gamma \vdash t : T$ holds, $\vdash \Gamma$ and $\Gamma \vdash T$ do too. This is enforced at the leaves of derivation trees, with rules like the following one for variables:

$$\text{wfVar} \frac{\vdash \Gamma \quad (x : T \in \Gamma)}{\Gamma \vdash x : T}$$

While this is sensible for a specification, it would be algorithmically much too costly, as it would mean re-checking the whole context at every leaf of the derivation. Instead, implementations maintain context well-formation as an invariant, rather than enforcing it: contexts are only ever extended with types which are known to be well-formed, but never fully checked.

More generally, algorithmic judgements have three kinds of arguments: inputs, outputs and a subject. From the algorithmic point of view, both inputs and the subject are arguments of the function. The difference lies in what we assume about them: inputs are assumed to be well-formed before calling the function, while it is the function's role to ensure the subject is valid. For instance, in type inference $\Gamma \vdash t \rhd T$, $\Gamma$ is an input, $t$ is the subject, and $T$ is an output. The invariant to maintain is twofold. First, when a function is called, its inputs must already be known to be well-formed. Second, when it returns positively, both its subject and output must also be well-formed. This idea, originally due to McBride [2018], and later dubbed "McBride discipline" in Lennon-Bertrand [2021, 2022], imposes clear constraints on how to design sensible bidirectional typing rules.

Most properties of algorithmic judgements only hold when their inputs are well-formed, and thus the invariant preservation must appear in proofs. In Lennon-Bertrand [2021], only soundness is shown directly for bidirectional typing, all other properties are shown on the declarative side and then transported to the bidirectional one. Thus, the McBride discipline is treated in an ad-hoc way, by carefully crafting the induction predicate for soundness to bake well-formation invariants in. Here, however, we need to show multiple properties of our algorithmic judgements, each needing their induction. The approach of modifying induction predicates would thus be inconvenient, because we would have to show the same invariant preservation over and over again in each inductive proof. Instead, in BundledAlgorithmicTyping we introduce *bundled* algorithmic judgement, and custom induction principles that handle invariant preservation once and for all.

Bundled algorithmic judgements, pack together an algorithmic judgement together with its pre-condition, *i.e.* well-formation of its inputs. These are the ones shown to be equivalent to the declarative ones. Note that only the "main" judgement is expressed algorithmically, all other ones are only declarative. This is because at this stage we have very few properties of algorithmic judgements, and it would thus be too difficult to construct these fields if they were expressed algorithmically. For declarative judgements, on the contrary, we can already rely on the declarative instance of the logical relation to give us powerful theorems.

While these bundled judgements are not inductively defined, they satisfy induction principles (BundledConvInduction and BundledTypingInduction). These trade a weaker conclusion – with extra well-formation hypotheses – for induction steps which are easier to prove, by having access to extra hypotheses. Concretely, in the induction step corresponding to infApp, on top of the induction hypotheses, one also knows that $\Gamma$ is well-typed (pre-condition to the conclusion), and that moreover $\Gamma \vdash f : \Pi\, x \colon A.B$, $\Gamma \vdash \Pi\, x \colon A.B$ and $\Gamma \vdash a : A$ (post-conditions of the premises).

We show these induction principles by regular induction on the unbundled algorithmic judgements. This amounts to showing once and for all invariant preservation, *i.e.* that pre-conditions to recursive calls are always satisfied provided pre-conditions of the conclusion and post-conditions of previous recursive calls. As a nice side-product, since part of the post-condition of the algorithmic judgements is their undirected counterparts, we get soundness of the bundled algorithmic judgements, for free.

### 6.4 Properties of algorithmic typing

Using bundled induction, we are able to show most properties of algorithmic conversion in AlgorithmicConvProperties. In particular, it is a partial equivalence relation, it is stable under conversion of the type (that is, if $\Gamma \vdash t \cong t' \triangleleft T$ and $\Gamma \vdash T \cong T'$, then $\Gamma \vdash t \cong t' \triangleleft T'$), and neutral comparison implies conversion, at any type.

Stability under conversion requires injectivity of type constructors: we must know that we can use the same rule for $T'$ that was used for $T$, so if *e.g.* $T$ reduces to a $\Pi$ type, then $T'$ should too, with recursively related domain and codomain to invoke induction hypotheses.

Inclusion of neutral comparison in conversion requires normalization. Indeed, if a type supports an $\eta$ rule, this rule is applied systematically. Hence, we can go from neutral to comparison only at types without extensionality rules. We thus need to know that this cannot go forever, *i.e.* that the type normalizes and so that it cannot produce $\Pi$ or $\Sigma$ type constructors forever.

Yet, we cannot show at this stage that bundled judgements form an instance of generic typing. To do so, we would need a mixed form of transitivity: if $\Gamma \vdash T \cong T'$ (algorithmically) and $\Gamma \vdash T' \cong T''$ (declaratively), then $\Gamma \vdash T \cong T''$ (algorithmically). The natural proof idea is that if the rule for algorithmic conversion is *e.g.* congruence of $\Pi$, then by injectivity also $T''$ must reduce to a $\Pi$-type since it is convertible to $T'$, and so the same rule applies. However, when we reach the base case of neutral types, we are stuck: as explained in Section 4.4, the logical relation does not show that neutrals are injective!

Instead, we first inhabit an "intermediate" instance of generic typing, where conversion is taken to be (bundled) algorithmic, but typing is declarative. For this instance we can show all the required properties at that stage. By the fundamental lemma, we obtain completeness of bundled algorithmic conversion. This is enough to establish properties of algorithmic typing (AlgorithmicTypingProperties) and inhabit a third, fully algorithmic instance, concluding that algorithmic typing is complete. Once we show that algorithmic typing is decidable (Section 7.2), we obtain that type checking is decidable for declarative typing.

## 7 Engineering aspects

We detail in this section some aspects of our implementation strategy that could inform future formal developments: first, the technical challenges to encode small induction recursion in CoQ; second, a report on our use of a recent library for manipulating partial functions; and third, the impact of tactics and automation on our development.

### 7.1 Encoding small induction-recursion, in practice

Section 5 explains the high-level strategy to implement a logical relation for MLTT using small induction-recursion. Here, we explain some implementation details that one must tackle in order to effectively encode small induction-recursion.

*Avoiding code duplication.* Compared to what is presented in Section 5, we generalize the type of LR to

```
LR@{i} : forall (ℓ : TypeLevel),
  (forall ℓ', ℓ' < ℓ → RedRel@{i}) → RedRel@{i+1}
```

where $i$ is quantified over by prenex universe polymorphism and the additional parameter is a form of open recursion call. Although in practice we only care about levels 0 and 1, we can define non-uniformly the logical relation for every closed numeral with a single inductive definition for LR.

This allows sharing the definition of the logical relation by packing our two definitions of reducibility into a universe polymorphic container, whose universe level does not depend on $\ell$ but can be instantiated to either level.

*Using the logical relation.* Once the logical relation is suitably encoded, we should avoid the specificities and intricacies of the encoding when proving properties. In particular, the induction principle automatically derived by CoQ is unsuitable. This is due to the wish to only ever manipulate the complete, packed presentation of the logical relation when proving properties, but also to nested inductive occurrences in the definition of the logical relation, for which CoQ lacks support when deriving induction principles. Thus, we show by hand a custom induction principle, which hides our encoding: users should never directly deal with it, but only interact with the logical relation through its encapsulation in the induction principle. In particular, the induction principle allows us to prove our lemmas on reducibility for all levels at once by making them universe polymorphic. Tactics also adapt to this custom induction principle, providing a proving experience very similar to what one would have with an induction recursion scheme supported directly by the proof assistant.

*Proving properties of the logical relation.* The properties of the logical relation presented in Section 4.1 need to be proved in a precise order visible in the dependency graph of the corresponding files Appendix A. Some properties would a priori need a mutual definition at types and terms levels: for instance reflexivity of reducible conversion at type $\mathbf{Id}_A\ x\ y$ require reflexivity proofs of reducible conversion for the subterms $x$ and $y$. We cut these dependencies by throwing in additional data in the logical relation that turns out to be redundant once we proved all the properties. We need to be careful that this additional data preserve the irrelevance of the logical relation with respect to reducible conversion. Indeed, irrelevance turns out to be the most difficult property to establish. In particular, we generalize the statement of the irrelevance of the logical relation so that it also encompasses cumulativity with respect to universe levels, and symmetry and transitivity of type conversion.

### 7.2 An executable type checker, in Coq

***Open recursion.*** The idea of our implementation, based on Winterhalter [2023], is to describe the calling graph of our algorithm, in the "open recursion" fashion. That way, we can use different approaches to execution, depending on our goals: a fuelled version can be efficiently run inside Coq, since it does not need to compute proofs; while a graph-based version leads to a certified total type checking function check, albeit one which should not be run without proof erasure. Moreover, this approach allows defining functions before arguing about their termination, which means we can run a certified sound type checker returning valid type derivations without yet having proven its termination. [5]

Concretely, Winterhalter [2023] provides a type `orec A B C` of computations returning values of type `C` using open recursive call of type `∀ (x : A), B x`. This type also allows undefined computations and calls to previously defined partial functions which is crucial to scale to a library consisting of multiple functions. As part of our development, we contributed a small extension to this calling mechanism that lower the required universe levels in the definition at the cost of making explicit the functions that can be called. The type `∇ x : A. B`, defined as `∀ (x : A), orec A B (B x)`, represents open dependent partial functions from `A` to `B`.

***Defining the partial functions.*** To define our partial functions in the most streamlined way, and to support effective reasoning about them, we rely on a wealth of techniques. The type `orec A B` describes the open recursion monad supporting the generic operations `rec`, `call` and `undefined`. However, some of the functions we want to use also naturally involve an exception monad, for instance type inference should return either a type or an error. This means that to write proper monadic code, we must combine the exception and open recursion monad, and allow to call a function using only one of the two monadic structure (reduction, which uses only open recursion monad but cannot error, or context access, which can error but is structurally recursive) inside the combined monad.

---
[5]This is visible in the dependency graph in Appendix A, where Functions only depends on the AST of terms, and Soundness only depends on the definition of AlgorithmicTyping, but not on the logical relation.

A nice aspect of the open recursion approach is its natural support for modularity. Because we do not close the recursion loop immediately, we are free to separately describe mutually defined functions. For instance, conversion checking corresponds to six mutually-defined routines, corresponding to the six conv_state tags. We can define separately the six corresponding functions (conv_ty, conv_ne, etc.), before finally combining them in the conv function. While we do this manually, encoding mutual recursion by non-mutual recursion with a tag, it might be interesting to explore the possibility to integrate such a mechanism directly into Winterhalter [2023].

The implementations themselves follow closely the algorithmic judgements. Only reduction is more complex: we implement a stack machine to be able to handle reduction under weak-head contexts. This machine is in essence a simplification of MetaCoq's. However, since we do not need to show its termination while defining it, we avoid the very complex dependent lexicographic order modulo a relation needed there [Winterhalter 2020, Chapters 21-23].

***Reasoning on partial functions.*** We wish to establish three properties of our partial checking functions, namely
- *soundness*: if the function returns without raising an error, then the corresponding judgement is derivable;
- *completeness*: if the corresponding judgement is derivable, then the function always returns a positive result;
- *termination*: the function always returns a result (either a positive one or an error) when called on inputs satisfying its precondition.

To prove Soundness, we use *functional induction*, induction principles tailored to the functions' calling graphs. A nice aspect of open recursion is that it makes it easy to perform this generically: a functional induction principle, which we can use out of the box, is part of Winterhalter [2023]. Since our functions directly follow the structure of the algorithmic judgements, establishing their soundness is straightforward, taking only a few lines of Ltac code – the difficult work is soundness of the algorithmic judgements with respect to the declarative ones. The only slightly subtle proof is for reduction, because of the stack machine used in the implementation.

Completeness is proven by bundled induction on algorithmic typing. Again, the main subtlety is in dealing with reduction. Indeed, because reduction cannot error, its completeness already encompasses termination, and we need to rely on a somewhat complex order, whose well-foundedness is established by normalization. Moreover, we need to show that whenever reduction throws an undefined error we are indeed in an unreachable branch. This means reasoning on the structure of well-typed stacks, to show that these branches correspond to ill-typed terms. Consequently, reduction is only complete when called on well-typed terms, and accordingly we need bundled induction to have this invariant available whenever reduction is called.

The last challenge is Termination of the conversion checking algorithm. We show that whenever $\Gamma \vdash t \cong u \triangleleft T$, then conversion checking terminates on the inputs $\Gamma$, $t$, $u'$ and $T$, for *any* well-typed $u'$ (unrelated with $u$). In essence, the structure of the proof of $\Gamma \vdash t \cong u \triangleleft T$ implicitly contains a derivation of deep normalization of $t$, including the relevant $\eta$-expansions, and we induct on that structure. Then, by reflexivity we know that any well-typed term $t$ is convertible to itself, which provides the derivation we need. Termination of type-checking is easy, since it is structural.

***Executing partial functions.*** From our functions, we can derive two implementations, with different purposes. The first is a fuelled implementation, which computes efficiently by induction on its extra natural number argument. Some examples are included in Execution, where we use the fuelled checker and its soundness to derive typing derivations by reflexion. Note how this only relies on soundness of the functions, but not on their completeness or termination. The second is the total implementation, which lets us derive a type checker with the type one expects for a proper decision procedure, in Decidability.

### 7.3 Automation and its limitations

***AutoSubst.*** We rely on the OCaml implementation of AutoSubst 2 [Stark et al. 2019; Dapprich 2021] to deal with all the aspect of raw syntax, define untyped renamings and substitutions, generating boilerplate lemmas for these, and provide tactical support to discharge equational obligations. We heavily use the auto-generated lemmas typically through the `asimpl` tactic, a decision procedure for equations in the substitution calculus, which greatly alleviate the burden of these tedious goals. Still, there is room for improving AutoSubst and its use.

First, on top of raw renamings (functions `nat → term`), we use an inductive notion of well-typed weakenings. Such a weakening can be turned into a renaming, but the cohabitation of the two notions makes the development cumbersome, and forces us to redefine AutoSubst's built-in `asimpl` to a tactic `bsimpl` which deals with this discrepancy. This problem partly stems from our formalization choices, rather than purely from AutoSubst, but it was unclear to us how to seamlessly combine inductive reasoning on weakenings with AutoSubst's renamings.

Second, the `asimpl`/`bsimpl` tactics, while providing useful decision procedures, have some practical limitations. They rely on **`setoid_rewrite`**, which quickly becomes slow, even on the not so large goals we have to handle, and become the performance bottleneck in a number of proof scripts. Moreover, they are only able to work on equations without existential variables (evars). This somewhat negates the otherwise powerful mechanism of evars, which lets one avoid giving a value upfront, rather refining it as one goes along with the proof.

***Type-classes and tags.*** As mentioned in Section 4.2, our logical relation is defined over a generic family of typing (and conversion, reduction, etc.) judgements. This is not only theoretically useful, but also doubles as a notational device, as we attach notations to the type classes for the generic judgements, in the Math Classes style [Spitters and Van Der Weegen 2011].

However, disambiguation between the different judgement families cannot be type-directed: all our typing judgements have the same type! Instead, we rely on a system of tags, inhabitants of a distinguished type `tag` with a single constructor `mkTag`. Each time we introduce a new family of judgements we also introduce an associated new opaque inhabitant of `tag`, and crucially keep the instances wrapped in a module. When working on this specific instance, this module can be imported, bringing the required instance in context. Similarly, when working generically, as usual with type classes, we introduce a hypothetical instance and tag. If there is only one specific or generic instance in context, it is safe to use the unqualified notation for typing `Γ ⊢ t : T`, which will find the unique instance available. If, however, multiple instances are available (typically, when working with both the declarative and algorithmic systems), we use a different notation with an explicit tag `Γ ⊢[de] t : T`, which will find the only instance with the corresponding tag, here the declarative one. This strategy is similar to that used by Allamigeon et al. [2023], to solve a similar disambiguation problem in the context of canonical structures.

***Automation.*** We use tactics to provide for judgement-independent notions, *e.g.* we use a single `irrelevance` tactic to use lemmas stating that one of the logical relation judgement is irrelevant in some of its parameters, or a `boundary` tactic to obtain "boundary" conditions of judgement, for instance to deduce from `Γ ⊢ t : T` that also `Γ ⊢ T`. An important part of the work achieved by the definition of the logical relation consist in its generalization of typing contexts through Kripke-style quantifications over renamings and substitutions, and we use instantiation tactics to automatically apply lemmas to the relevant hypotheses. Finally, to handle goals easily solved by using properties of a generic typing judgements, we provide `gen_typing` tactic, which performs a crude proof search. While these tactics already go some way in making proof writing higher level and more robust, we feel like there is a lot of space for designing tactics which are more powerful, robust and predictable. A typical issue with `gen_typing`, for instance, is that it either succeeds quickly or fails excruciatingly slowly, resulting in brittleness in the face of proof changes.

## 8 Future work

### 8.1 Extensions and improvements

*Universes.* It should be possible to add more universes to obtain a hierarchy of arbitrary finite length, and we see no theoretical obstacle in doing so, although there might be some practical metaprogramming challenges in the definition of the logical relation. Going beyond this and tackling a full countable hierarchy requires a different approach. If we want to stay in axiom-free Coq and keep avoiding induction-recursion, the natural candidate is to use impredicativity. However, in Coq the impredicative sort `Prop` comes with restrictions, and it is currently unclear to us whether these restrictions would break a naïve port of our development to a logical relation in `Prop`.

*Inductive types.* Our current formalization only handles rather simple inductive types, namely **N**, and **Id**. While these already encompass the main difficulties posed by inductive types, a natural extension would be to add W types: together with **Id**, Σ, Π and a few base types, these can encode all indexed inductive types [Hugunin 2020; Awodey et al. 2012], which would really narrow the gap between our object and metatheory to a difference in universes. A more ambitious step would be to consider a general inductive scheme, as used in virtually any realistic system, and in MetaCoq, instead of particular examples.

*Less naïve algorithms.* Our current algorithm is a naïve one, closely following the logical relation. An interesting improvement would be to implement term-directed extensionality rules [Abel and Coquand 2007]. Lennon-Bertrand [2022, Chapter 6] shows that once one has access to the meta-theory of the unoptimized, algorithmic variant, the proof of equivalence is straightforward. On the bidirectional side, we mentioned in Section 6 the common pattern, used for instance in the kernel of Agda [Norell 2007], of trading lighter annotations – typically, unannotated abstraction – against incomplete inference – some terms only check. While this makes type inference in general incomplete, type checking should stay decidable in a suitable sense. We believe that adapting our formalization to that setting is not only possible, but would be a very interesting endeavour.

*Automation.* On the practical side, we feel like there is a lot of room to improve automation, taking inspiration from the rich Coq ecosystem. Indeed, the main difficulty for a proof by logical relations is in the setup of the relation, but most proof obligations are rather repetitive and unsurprising. While our tactics already relieve us from quite a bit of this tedious work, they are far from alleviating all the pain.

*Integration in MetaCoq.* While there is no formal relation between the present work and MetaCoq, we hope that in the future we will be able to connect the two, showing that the normalization axiom of MetaCoq is provable at least for a subset of the language. This is challenging, because there is still a significant gap between the two systems: we use typed conversion while MetaCoq's is untyped, MetaCoq uses pattern-matching and (guarded) fixpoints instead of recursors. The techniques we deploy to implement our type checker in Section 7.2 should be useful in MetaCoq too, delimiting further the portion of the code that depends on normalization.

### 8.2 Applications

Although our development is centered on MLTT, its modularity makes it amenable to study other type theories.

*Definitional functor laws.* An ongoing parallel project building on this work extends MLTT with definitional functor laws for the `map` operation on lists: `map id l ≡ l` and `map (f o g) l ≡ map f (map g l)`. In particular, the proofs of normalization and decidability adapt with relatively little changes on the original formalization.

*Strict propositions and* TT[obs]. Pujet and Tabareau [2023] were the first to attempt removing induction-recursion from their normalization proof, in order to provide a conservativity result for their theory TT[obs]: every numeric function that is definable in it can also be defined in MLTT. Yet, they still rely on first-class universe levels, a feature of Agda with little theoretical investigation. Moreover, they cannot restrict Agda's positivity checker to only allow for "standard" inductive definitions, and as a matter of fact we had to significantly amend their inductively defined logical relation to have it accepted by Coq. Thus, it would be natural to solidify the conservativity result of Pujet and Tabareau [2023] by porting their development of TT[obs] to our setting.

*Variants of* MLTT *and the multiverse.* An enticing potential application of this work, beyond the implementation of various extension, is the ability to explore the interactions of multiple extensions. This could take the shape of adding multiple universes in order to delimit potentially incompatible extensions, e.g. with uniqueness of identity proofs such as TT[obs] and univalence such as some variant of homotopy type theory or cubical type theories, and study which interactions are sound, in the sense that they preserve the metatheorems established in this formalization.

## References

Andreas Abel. 2010. "Towards Normalization by Evaluation for the $\beta\eta$-Calculus of Constructions." In: *Functional and Logic Programming, 10th*

*International Symposium, FLOPS 2010, Sendai, Japan, April 19-21, 2010. Proceedings* (Lecture Notes in Computer Science). Ed. by Matthias Blume, Naoki Kobayashi, and Germán Vidal. Vol. 6009. Springer-Verlag, 224–239. ISBN: 978-3-642-12250-7.

Andreas Abel, Klaus Aehlig, and Peter Dybjer. 2007. "Normalization by Evaluation for Martin-Löf Type Theory with One Universe." In: *Proceedings of the 23rd Conference on the Mathematical Foundations of Programming Semantics (MFPS XXIII), New Orleans, LA, USA, 11-14 April 2007* (Electronic Notes in Theoretical Computer Science). Ed. by Marcelo Fiore. Vol. 173. Elsevier, 17–39.

Andreas Abel, Guillaume Allais, Aliya Hameer, Brigitte Pientka, Alberto Momigliano, Steven Schäfer, and Kathrin Stark. 2019. "POPLMark reloaded: Mechanizing proofs by logical relations." *Journal of Functional Programming*, 29, e19. DOI: 10.1017/S0956796819000170.

Andreas Abel and Thierry Coquand. 2007. "Untyped Algorithmic Equality for Martin-Löf's Logical Framework with Surjective Pairs." *Fundamenta Informaticae*, 77, 4, 345–395. TLCA'05 special issue. http://fi.mimuw.edu.pl/abs77.html#15.

Andreas Abel, Thierry Coquand, and Peter Dybjer. 2007. "Normalization by Evaluation for Martin-Löf Type Theory with Typed Equality Judgements." In: *22nd IEEE Symposium on Logic in Computer Science (LICS 2007), 10-12 July 2007, Wroclaw, Poland, Proceedings*. "IEEE Computer Society Press", 3–12. DOI: 10.1109/LICS.2007.33.

Andreas Abel, Thierry Coquand, and Miguel Pagano. 2009. "A Modular Type-Checking Algorithm for Type Theory with Singleton Types and Proof Irrelevance." In: *Typed Lambda Calculi and Applications*. Ed. by Pierre-Louis Curien. Springer Berlin Heidelberg, Berlin, Heidelberg, 5–19.

Andreas Abel, Nils Anders Danielsson, and Oskar Eriksson. Aug. 2023. "A Graded Modal Dependent Type Theory with a Universe and Erasure, Formalized." *Proc. ACM Program. Lang.*, 7, ICFP, Article 220, (Aug. 2023), 35 pages. DOI: 10.1145/3607862.

Andreas Abel, Joakim Öhman, and Andrea Vezzosi. Dec. 2017. "Decidability of Conversion for Type Theory in Type Theory." *Proc. ACM Program. Lang.*, 2, POPL, Article 23, (Dec. 2017), 29 pages. DOI: 10.1145/3158111.

Oskar Abrahamsson, Magnus O. Myreen, Ramana Kumar, and Thomas Sewell. 2022. "Candle: A Verified Implementation of HOL Light." In: *13th International Conference on Interactive Theorem Proving (ITP 2022)* (Leibniz International Proceedings in Informatics (LIPIcs)). Ed. by June Andronick and Leonardo de Moura. Vol. 237. Schloss Dagstuhl – Leibniz-Zentrum für Informatik, Dagstuhl, Germany, 3:1–3:17. ISBN: 978-3-95977-252-5. DOI: 10.4230/LIPIcs.ITP.2022.3.

Arthur Adjedj, Meven Lennon-Bertrand, Kenji Maillard, Pierre-Marie Pédrot, and Loïc Pujet. 2023a. *Logical Relation for MLTT in Coq*. https://github.com/CoqHott/logrel-coq. (2023).

[SW] Arthur Adjedj, Meven Lennon-Bertrand, Kenji Maillard, Pierre-Marie Pédrot, and Loïc Pujet, *Martin-Löf à la Coq* version cpp24-submission, Sept. 2023. DOI: 10.5281/zenodo.8367154, URL: https://doi.org/10.5281/zenodo.8367154.

Xavier Allamigeon, Quentin Canu, Cyril Cohen, Kazuhiko Sakaguchi, and Pierre-Yves Strub. 2023. "Design patterns of hierarchies for order structures." working paper or preprint. (2023). https://hal.inria.fr/hal-04008820.

Steve Awodey, Nicola Gambino, and Kristina Sojakova. 2012. "Inductive Types in Homotopy Type Theory." In: *Proceedings of the 2012 27th Annual IEEE/ACM Symposium on Logic in Computer Science* (LICS '12). IEEE Computer Society, 95–104. ISBN: 9780769547695. DOI: 10.1109/LICS.2012.21.

Brian Aydemir et al.. 2005. "Mechanized metatheory for the masses: the POPLmark challenge." In: *International Conference on Theorem Proving in Higher Order Logics*. Springer, 50–65.

Bruno Barras. 1999. "Auto-validation d'un système de preuves avec familles inductives." Ph.D. Dissertation.

Bruno Barras. 2014. "Semantical Investigations in Intuitionistic Set Theory and Type Theories with Inductive Families." Habilitation thesis. (2014). http://www.lsv.fr/~barras/habilitation/.

Bruno Barras and Benjamin Werner. 1997. "Coq in Coq." (1997). http://www.lix.polytechnique.fr/Labo/Bruno.Barras/publi/coqincoq.pdf.

Rafaël Bocquet, Ambrus Kaposi, and Christian Sattler. 2023. *For the Metatheory of Type Theory, Internal Sconing Is Enough*. (2023). arXiv: 2302.05190 [cs.LO].

Ana Bove and Venanzio Capretta. 2005. "Modelling general recursion in type theory." *Mathematical Structures in Computer Science*, 15, 4, 671–708. DOI: 10.1017/S0960129505004822.

Venanzio Capretta. July 2005. "General Recursion via Coinductive Types." *Logical Methods in Computer Science*, Volume 1, Issue 2, (July 2005). DOI: 10.2168/LMCS-1(2:1)2005.

Thierry Coquand. 2018. "Canonicity and normalisation for Dependent Type Theory." *CoRR*, abs/1810.09367. arXiv: 1810.09367.

Adrian Dapprich. 2021. "Generating Infrastructural Code for Terms with Binders using MetaCoq and OCaml." Bachelor Thesis. Saarland University.

Benjamin Delaware, Bruno C. d. S. Oliveira, and Tom Schrijvers. 2013. "Meta-theory à la carte." In: *The 40th Annual ACM SIGPLAN-SIGACT Symposium on Principles of Programming Languages, POPL '13, Rome, Italy - January 23 - 25, 2013*. Ed. by Roberto Giacobazzi and Radhia Cousot. ACM, 207–218. ISBN: 978-1-4503-1832-7. DOI: 10.1145/2429069.2429094.

Gilles Dowek. 1993. "The undecidability of typability in the Lambda-Pi-calculus." In: *Typed Lambda Calculi and Applications*. Ed. by Marc Bezem and Jan Friso Groote. Springer Berlin Heidelberg, Berlin, Heidelberg, 139–145.

Jana Dunfield and Neel Krishnaswami. May 2021. "Bidirectional Typing." *ACM Computing Surveys*, 54, 5, Article 98, (May 2021), 38 pages. DOI: 10.1145/3450952.

Peter Dybjer and Anton Setzer. 2003. "Induction-recursion and initial algebras." *Annals of Pure and Applied Logic*, 124, 1-3, 1–47. DOI: 10.1016/S0168-0072(02)00096-9.

Yannick Forster and Kathrin Stark. 2020. "Coq à la carte: a practical approach to modular syntax with binders." In: *Proceedings of the 9th ACM SIGPLAN International Conference on Certified Programs and Proofs, CPP 2020, New Orleans, LA, USA, January 20-21, 2020*. Ed. by Jasmin Blanchette and Catalin Hritcu. ACM, 186–200. ISBN: 978-1-4503-7097-4. DOI: 10.1145/3372885.3373817.

Gaëtan Gilbert, Jesper Cockx, Matthieu Sozeau, and Nicolas Tabareau. Jan. 2019. "Definitional Proof-Irrelevance without K." *Proceedings of the ACM on Programming Languages*. POPL'19 3, POPL, (Jan. 2019), 1–28. DOI: 10.1145/329031610.1145/3290316.

Jean-Yves Girard, Paul Taylor, and Yves Lafont. Apr. 1989. *Proofs and Types*. Cambridge University Press, (Apr. 1989). ISBN: 0521371813.

Daniel Gratzer, Jonathan Sterling, and Lars Birkedal. July 2019. "Implementing a Modal Dependent Type Theory." *Proc. ACM Program. Lang.*, 3, ICFP, Article 107, (July 2019), 29 pages. DOI: 10.1145/3341711.

Peter Hancock, Conor McBride, Neil Ghani, Lorenzo Malatesta, and Thorsten Altenkirch. 2013. "Small Induction Recursion." In: *Typed Lambda Calculi and Applications*. Ed. by Masahito Hasegawa. Springer Berlin Heidelberg, Berlin, Heidelberg, 156–172.

Jasper Hugunin. 2020. "Why Not W?" In: *26th International Conference on Types for Proofs and Programs, TYPES 2020, March 2-5, 2020, University of Turin, Italy* (LIPIcs). Ed. by Ugo de'Liguoro, Stefano Berardi, and Thorsten Altenkirch. Vol. 188. Schloss Dagstuhl - Leibniz-Zentrum für Informatik, 8:1–8:9. ISBN: 978-3-95977-182-5. DOI: 10.4230/LIPIcs.TYPES.2020.8.

Ende Jin, Nada Amin, and Yizhou Zhang. June 2023. "Extensible Metatheory Mechanization via Family Polymorphism." *Proc. ACM Program. Lang.*, 7, PLDI, Article 172, (June 2023), 25 pages. DOI: 10.1145/3591286.

András Kovács. 2022. "Generalized Universe Hierarchies and First-Class Universe Levels." In: *30th EACSL Annual Conference on Computer Science Logic (CSL 2022)* (Leibniz International Proceedings in Informatics (LIPIcs)). Ed. by Florin Manea and Alex Simpson. Vol. 216. Schloss Dagstuhl – Leibniz-Zentrum für Informatik, Dagstuhl, Germany, 28:1–28:17. ISBN: 978-3-95977-218-1. DOI: 10.4230/LIPIcs.CSL.2022.28.

Dominique Larchey-Wendling and Jean-François Monin. 2022. "The Braga method: Extracting certified algorithms from complex recursive schemes in Coq." In: *PROOF AND COMPUTATION II: From Proof Theory and Univalent Mathematics to Program Extraction and Verification*. World Scientific, 305–386.

Meven Lennon-Bertrand. 2022. "Bidirectional Typing for the Calculus of Inductive Constructions." Ph.D. Dissertation. Nantes Université.

Meven Lennon-Bertrand. 2021. "Complete Bidirectional Typing for the Calculus of Inductive Constructions." In: *12th International Conference on Interactive Theorem Proving (ITP 2021)* (Leibniz International Proceedings in Informatics (LIPIcs)). Ed. by Liron Cohen and Cezary Kaliszyk. Vol. 193. Schloss Dagstuhl – Leibniz-Zentrum für Informatik. ISBN: 978-3-95977-188-7. DOI: 10.4230/LIPIcs.ITP.2021.24.

Xavier Leroy. Feb. 13, 2020. *Coq en Coq*. slides https://xavierleroy.org/CdF/2019-2020/8.pdf; Literature review between 50' and 60'. (Feb. 13, 2020). https://youtu.be/nuLJ5S9qh-I.

Per Martin-Löf and Giovanni Sambin. 1984. *Intuitionistic Type Theory*. Studies in Proof Theory 1. Napoli: Bibliopolis.

Conor McBride. Aug. 6, 2018. "Basics of Bidirectionalism." Blog post. (Aug. 6, 2018). https://pigworker.wordpress.com/2018/08/06/basics-of-bidirectionalism/.

Conor McBride. 2015. "Turing-Completeness Totally Free." In: *Mathematics of Program Construction*. Ed. by Ralf Hinze and Janis Voigtländer. Springer International Publishing, Cham, 257–275.

Ulf Norell. Sept. 2007. "Towards a practical programming language based on dependent type theory." Ph.D. Dissertation. Department of Computer Science and Engineering, Chalmers University of Technology, (Sept. 2007).

Anja Petković Komel. 2021. "Meta-analysis of type theories with an application to the design of formal proofs." Ph.D. Dissertation. University of Ljubljana.

Benjamin C. Pierce and David N. Turner. Jan. 2000. "Local Type Inference." *ACM Transactions on Programming Languages and Systems*, 22, 1, (Jan. 2000), 1–44. DOI: 10.1145/345099.345100.

Loïc Pujet and Nicolas Tabareau. Jan. 2023. "Impredicative Observational Equality." *Proc. ACM Program. Lang.*, 7, POPL, Article 74, (Jan. 2023), 26 pages. DOI: 10.1145/3571739.

Loïc Pujet and Nicolas Tabareau. 2022. "Observational Equality: Now for Good." *Proc. ACM Program. Lang.*, 6, POPL, Article 32, 27 pages. DOI: 10.1145/3498693.

Morten Heine Sørensen and Pawel Urzyczyn. 2006. *Lectures on the Curry-Howard isomorphism*. Studies in Logic and the Foundations of Mathematics. Vol. 149. Elsevier Science.

Matthieu Sozeau, Abhishek Anand, Simon Boulier, Cyril Cohen, Yannick Forster, Fabian Kunze, Gregory Malecha, Nicolas Tabareau, and Théo Winterhalter. Feb. 2020. "The MetaCoq Project." *Journal of Automated Reasoning*, (Feb. 2020). DOI: 10.1007/s10817-019-09540-0.

Matthieu Sozeau, Yannick Forster, Meven Lennon-Bertrand, Jakob Botsch Nielsen, Nicolas Tabareau, and Théo Winterhalter. Apr. 2023. "Correct and Complete Type Checking and Certified Erasure for Coq, in Coq." Preprint. (Apr. 2023). https://inria.hal.science/hal-04077552.

Bas Spitters and Eelis Van Der Weegen. 2011. "Type classes for mathematics in type theory." *Mathematical Structures in Computer Science*, 21, 4, 795–825. DOI: 10.1017/S0960129511000119.

Kathrin Stark, Steven Schäfer, and Jonas Kaiser. 2019. "Autosubst 2: reasoning with multi-sorted de Bruijn terms and vector substitutions." In: *Proceedings of the 8th ACM SIGPLAN International Conference on Certified Programs and Proofs, CPP 2019, Cascais, Portugal, January 14-15, 2019*. Ed. by Assia Mahboubi and Magnus O. Myreen. ACM, 166–180. ISBN: 978-1-4503-6222-1. DOI: 10.1145/3293880.3294101.

Jonathan Sterling. Nov. 2021. "First Steps in Synthetic Tait Computability: The Objective Metatheory of Cubical Type Theory." Ph.D. Dissertation. Carnegie Mellon University, (Nov. 2021). DOI: 10.5281/zenodo.6990769. Doctoral thesis of Jonathan Sterling, Carnegie Mellon University.

Qian Wang and Bruno Barras. 2013. "Semantics of Intensional Type Theory extended with Decidable Equational Theories." In: *Computer Science Logic 2013 (CSL 2013), CSL 2013, September 2-5, 2013, Torino, Italy* (LIPIcs). Ed. by Simona Ronchi Della Rocca. Vol. 23. Schloss Dagstuhl - Leibniz-Zentrum für Informatik, 653–667. ISBN: 978-3-939897-60-6. DOI: 10.4230/LIPIcs.CSL.2013.653.

Paweł Wieczorek and Dariusz Biernacki. 2018. "A Coq Formalization of Normalization by Evaluation for Martin-Löf Type Theory." In: *Proceedings of the 7th ACM SIGPLAN International Conference on Certified Programs and Proofs* (CPP 2018). Association for Computing Machinery, Los Angeles, CA, USA, 266–279. ISBN: 9781450355865. DOI: 10.1145/3167091.

Théo Winterhalter. 2023. "Composable partial functions in Coq, totally for free." In: *29th International Conference on Types for Proofs and Programs*.

Théo Winterhalter. 2020. "Formalisation and meta-theory of type theory." Ph.D. Dissertation. Université de Nantes.

# A  Dependency graph of the library

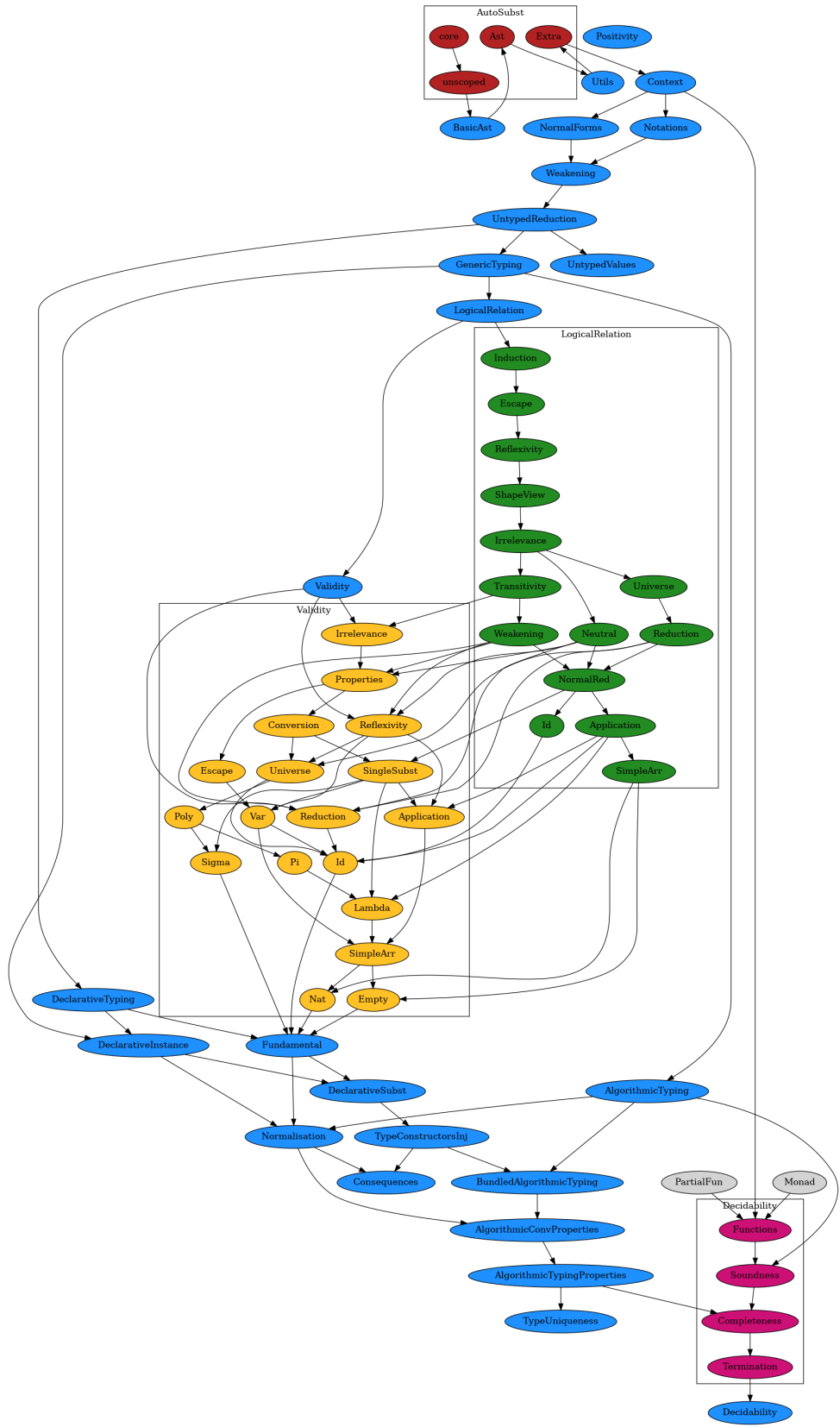